\documentclass[final,3p,times,twocolumn]{elsarticle}

\usepackage{graphicx}
\usepackage{braket}
\usepackage{bm}
\usepackage{amsfonts}
\usepackage{amsmath}
\usepackage{amssymb}

\begin{document}

\begin{frontmatter}

\title{Antiferromagnetic Spin Chain Behavior and a Transition to 3D Magnetic Order in Cu(\texttt{D},\texttt{L}-alanine)$_2$: Roles of H-bonds}


\author[1]{Rafael Calvo}
\author[1]{Rosana P. Sartoris}
\author[2]{Hern\'an L. Calvo}
\author[3,4]{Edson F. Chagas}
\author[3]{Raul E. Rapp}

\address[1]{Departamento de F\'isica, Facultad de Bioqu\'imica y Ciencias Biol\'ogicas, Universidad Nacional del
Litoral, and Instituto de F\'isica del Litoral (UNL--CONICET), G\"uemes 3450, 3000 Santa Fe, Argentina}
\address[2]{Instituto de F\'isica Enrique Gaviola (IFEG--CONICET) and FaMAF, Universidad Nacional de C\'ordoba, Ciudad Universitaria, 5000 
C\'ordoba, Argentina}
\address[3]{Instituto de F\'isica, Universidade Federal do Rio de Janeiro, CP 68528, 21941-972, Rio de Janeiro RJ, Brazil}
\address[4]{Instituto de F\'isica, Universidade Federal de Mato Grosso, 78060-900, Cuiab\'a-MT, Brazil}

\begin{keyword}
{Spin chains; Phase transition; Exchange interactions; H-bonds}
\end{keyword}


\begin{abstract}
We study the spin chain behavior, a transition to 3D magnetic order and the magnitudes of the exchange interactions for the metal-amino acid complex Cu(\texttt{D},\texttt{L}-alanine)$_2$$\cdot$H$_2$O, a model compound to investigate exchange couplings supported by chemical paths characteristic of biomolecules. Thermal and magnetic data were obtained as a function of temperature ($T$) and magnetic field ($B_0$). The magnetic contribution to the specific heat, measured between 0.48 and 30 K, displays above 1.8 K a 1D spin-chain behavior that can be fitted with an intrachain antiferromagnetic (AFM) exchange coupling constant $2J_0 = (-2.12 \pm 0.08)$ cm$^{-1}$ (defined as $\mathcal{H}_\text{ex}(i,i+1) = -2J_0 \bm{S}_i \cdot \bm{S}_{i+1}$), between neighbor coppers at 4.49 \AA{} along chains connected by non-covalent and H-bonds. We also observe a narrow specific heat peak at 0.89 K indicating a phase transition to a 3D magnetically ordered phase. Magnetization curves at fixed $T = 2$, 4 and 7 K with $B_0$ between 0 and 9 T, and at $T$ between 2 and 300 K with several fixed values of $B_0$ were globally fitted by an intrachain AFM exchange coupling constant $2J_0 = (-2.27 \pm 0.02)$ cm$^{-1}$ and $g = 2.091 \pm 0.005$. Interchain interactions $J_1$ between coppers in neighbor chains connected through long chemical paths with total length of 9.51 \AA{} cannot be estimated from magnetization curves. However, observation of the phase transition in the specific heat data allows estimating the range $0.1 \leqslant |2J_1| \leqslant 0.4$ cm$^{-1}$, covering the predictions of various approximations. We analyze the magnitudes of $2J_0$ and $2J_1$ in terms of the structure of the corresponding chemical paths. The main contribution in supporting the intrachain interaction is assigned to H-bonds while the interchain interactions are supported by paths containing H-bonds and carboxylate bridges, with the role of the H-bonds being predominant. We compare the obtained intrachain coupling with studies of compounds showing similar behavior and discuss the validity of the approximations allowing to calculate the interchain interactions.
\end{abstract}

\end{frontmatter}

\section{Introduction}

One-dimensional (1D) spin-chains provide insights on treating soluble quantum many-body problems, and became paradigmatic since the origins of quantum mechanics~\cite{Bethe1931, Mattis1993, Georges2001, Giamarchi2003}. The interest flourished when it was observed that chain compounds of metal ions with inorganic and organic ligands really exist in nature~\cite{Fritz1957, Haseda1961, DeJongh1974}, and that their thermodynamic properties can be modeled by the theories~\cite{Bonner1964, Griffiths1964, Griffiths1964a, Carlin1986, Kahn1993}. The compound copper tetrammine sulfate monohydrate, CTS = Cu(NH$_3$)$_4$SO$_4 \cdot$H$_2$O, was found to have 1D antiferromagnetic (AFM) behavior above 0.5 K~\cite{Fritz1957, Haseda1961}, that was attributed to Cu$^\text{II}$ ions located in chains parallel to the $\bm{a}$-axis~\cite{Morosin1969} linked as Cu$^\text{II}$---H$_2$O---Cu$^\text{II}$---H$_2$O---Cu$^\text{II}$, a result that started important theoretical developments that have also been helpful to deal with other problems~\cite{Mattis1993} such as molecular magnets~\cite{Georges2001, Kahn1993, Blundell2007}, superconductivity~\cite{Vuletic2006}, quantum dissipation~\cite{Werner2005}, and quantum communication~\cite{Bose2007}.

Mermin and Wagner~\cite{Mermin1966} proved that at any nonzero temperature ($T$), 1D or 2D isotropic spin Heisenberg model with finite-range exchange interaction can be neither FM nor AFM, and no transitions to ordered phases should occur by lowering $T$. However, in any real system, small interactions between chains (or layers) exist, introducing new ingredients to the problem. One indication of these interactions are phase transitions to long range magnetically ordered phases which have been observed in ``quasi'' 1D spin chains involving inorganic and organic ligands. For example, a transition arising from exchange couplings between Cu$^\text{II}$ in neighbor chains was observed~\cite{Haseda1961} in the specific heat of CTS at $T = 0.37$ K and attributed to interactions through weak $\cdots$Cu$^\text{II}$---NH$_3$---SO$_4$---NH$_3$---Cu$^\text{II}\cdots$ paths containing H-bonds. Further EPR experiments~\cite{Date1975, Duffy1977, Mollymoto1980} introduced doubts in the interpretation of the data about CTS, the compound that promoted the first theoretical advances about the thermodynamic properties of spin chains~\cite{Bonner1964, Griffiths1964}. In any case, this and other low-dimensional compounds provided a full source of information about thermodynamic properties and spin dynamics. The relationship between weak interchain interactions and phase transitions was studied experimentally~\cite{Haseda1961, DeJongh1974, Chagas2006}, and theoretically~\cite{Giamarchi2003, Oguchi1955, Ohmae1995, Matsumoto2000, Sakakibara2002, Oguchi1964, Smart1966, Schulz1996, Irkhin2000, Yasuda2005, Kokalj2015}, to evaluate these weak interactions and to verify the applicability of the subjacent theories.

The crystal and electronic structures of metal-amino acid and metal-peptide compounds provide model systems for metalloproteins in their active sites~\cite{Fleck2014}. The amino acid side chains offer an ample variety of chemical paths for exchange interactions between metal ions, and the compounds display appealing phenomena with one, two, or three-dimensional magnetic behavior and spin dynamics. Some long superexchange paths, characteristic of protein structures~\cite{Chagas2006, Calvo1993, Brondino1993, Costa-Filho2001, Santana2005}, are important because the magnitude of the exchange interactions between unpaired spins at long distances provide chemical information and allow estimating matrix elements for electron transfer processes between redox centers~\cite{Calvo2000, Calvo2001}. So, the exchange couplings in metal-amino acid complexes have been studied performing magnetic~\cite{Newman1976, Calvo1991, Calvo1999, Rizzi2003, Gerard2007, Calvo1991a}, thermal~\cite{Chagas2006, Calvo1982, Wakamatsu1989, Siqueira1993, Rapp1995}, EPR~\cite{Costa-Filho2001, Levstein1990, Levstein1990a, Calvo1991b, Martino1995, Martino1996, Costa-Filho2004, Neuman2014},
and NMR~\cite{Sandreczki1979, Szalontai2015} in Cu$^\text{II}$, Ni$^\text{II}$, Co$^\text{II}$ and Cr$^\text{III}$ compounds. In some cases it has been shown that the magnitudes of the weak interactions evaluated from thermodynamic data (magnetic or thermal) are similar to those obtained from EPR measurements, a technique allowing to separate the effects of interactions with widely different magnitudes (\textit{e.g.}, intrachain from interchain, or intralayer from interlayer), enabling to compare values obtained in different experiments~\cite{Chagas2006, Costa-Filho2001, Santana2005, Calvo2001, Calvo2007}. Unfortunately, this possibility strongly depends on characteristics of the compound and in many cases only the magnetic phase transitions at very low $T$~\cite{Chagas2006, Newman1976, Calvo1982, Wakamatsu1989, Rapp1995} contribute with new information about weak couplings supported by long non covalent chemical paths. Thus, specific heat measurements are particularly appropriate for this purpose~\cite{Matsumoto2000, Sakakibara2002}.

The crystal structure reported for Cu(\texttt{D},\texttt{L}-alanine)$_2$ monohydrate~\cite{Calvo1991, Hitchman1987} [Cu(NH$_2$CHCH$_3$CO$_2$)$_2\cdot$H$_2$O], henceforth called Cu(\texttt{D},\texttt{L}-ala)$_2$ shows Cu$^\text{II}$ ions in chains connected by water oxygen apical ligands common to neighbor coppers in the chains (see Fig.~\ref{fig:1}a). Static magnetic susceptibility data~\cite{Calvo1991} indicated a spin chain behavior with AFM exchange coupling $2J_0 = -2.2$ cm$^{-1}$ ($\mathcal{H}_\text{ex} = -\sum_i 2J_0 \bm{S}_i \cdot \bm{S}_{i+1}$). Single crystal EPR measurements at X-band showed a single resonance for all orientations of the magnetic field ($B_0 = \mu_0 H$, where $\mu_0$ is the vacuum permeability) arising from the collapse of the resonances~\cite{Anderson1954} corresponding to two rotated copper ions in the chains, produced by a relatively large intrachain exchange interaction. Here we report specific heat measurements between 0.48 and 30 K, and magnetization measurements between 2 and 90 K with applied \textit{dc} fields $0 \leqslant B_0 \leqslant 9$ T, in Cu(\texttt{D},\texttt{L}-ala)$_2$. The specific heat data display the broad peak expected for a spin chain approaching short range order~\cite{Bonner1964, Carlin1986, Kahn1993}, and also a narrow peak at lower $T$, which is attributed to a transition to a 3D magnetically ordered phase. These results reflect the linear chain behavior and also the presence of weak interchain exchange couplings between spins in neighbor chains producing the phase transition at low $T$, that are estimated under different approximations. The evaluated intra- and interchain exchange couplings are discussed in terms of the structures of the paths connecting the metal ions, and compared with results in other compounds. Our results emphasize the role of H-bonds to support both intrachain and interchain exchange couplings.

\section{Experimental procedures} 

Basic copper carbonate (0.5 mmol) was added to a solution of the racemic mixture \texttt{D},\texttt{L}-alanine (2 mmol in 20 mL of hot water). After kept at 60$^{\circ}$C until the reaction was complete, the remaining insoluble copper carbonate was filtered out with Millipore membranes having pore size 0.22 $\mu$m. Single crystals of good quality with sizes up to 4$\times$2$\times$2 mm$^3$ were obtained in about two weeks by slow evaporation of the solution at 40$^{\circ}$C. The compound is stable, allowing accurate magnetic measurements.

The heat capacity was measured in the $T$ range $0.48 < T < 30$ K with a semi-adiabatic calorimeter mounted in a pumped $^3$He cryostat. The polycrystalline sample was prepared as to improve the low thermal conductivity of the organic material at low $T$~\cite{Siqueira1993, Rapp1995, Torikachvili1983}. About 100 mg of powdered crystals of Cu(\texttt{D},\texttt{L}-ala)$_2$ with particle size of about 100 $\mu$m were mixed to about the same mass of powdered copper with particle size of about 5 $\mu$m, to obtain a good thermal conducting copper path around each sample particle, while keeping small the contribution of the metallic copper. The sample-copper mixture was compressed inside a cylindrical holder made of 0.04 mm thick copper foil, forming a chip, and allowing obtaining a temperature equilibration time constant $<25$ seconds within the sample, in the full measuring range. This chip was glued with vacuum grease to a copper plate fixed to the calorimeter, where the thermometer and heater were attached. This method allows fast replacement of samples in the calorimeter, and avoids destroying them after the measurement. Heat was applied during intervals of about 100 seconds as to produce $T$ changes of ~1\%. The calorimeter addendum heat capacity, always smaller than 1\% of that of the sample, was measured independently. The heat capacity of the sample, $C_\text{exp}$, was obtained subtracting the contribution of the copper (powder and foil) from the total heat capacity of the chip. The dispersion of the data was less than 2\%.

A commercial PPMS magnetometer (Quantum Design, Inc., San Diego, CA) was used to measure the magnetization using a powder sample of (56.4 $\pm$ 0.3) mg of Cu(\texttt{D},\texttt{L}-ala)$_2$. The contribution of the sample holder was measured separately, and subtracted from the results. Isothermal magnetization curves at $T = 2$, 4 and 7 K with $B_0$ up to 9 T, and magnetization curves with $T$ between ~2 and 90 K, for several applied fields $B_0$, were obtained. All reported magnetic measurements were corrected for diamagnetism using standard values estimated from Pascal constants and TIP~\cite{Carlin1986, Kahn1993}. The susceptibility $\chi_0(T)$ calculated from magnetization data obtained with $B_0 = 0.1$ T reproduce previous results~\cite{Calvo1991}.

\section{Crystal structure of Cu(\texttt{D},\texttt{L}-ala)$_2$ monohydrate}

\begin{figure}[tbp]
\begin{center}
\includegraphics[width=0.45\textwidth]{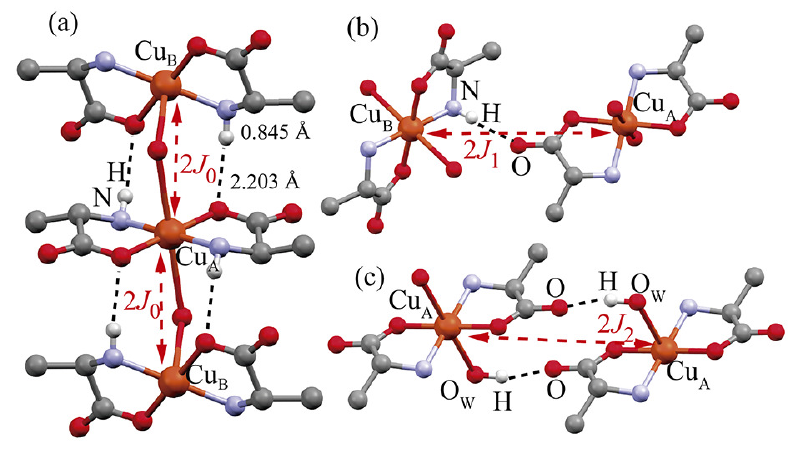}
\end{center}
\caption{Structure of Cu(\texttt{D},\texttt{L}-ala)$_2\cdot$H$_2$O according to Refs.~\cite{Calvo1991, Hitchman1987}. (a) View of the
copper chains along $\bm{c}$, emphasizing the individual molecules and the paths connecting copper neighbors in a chain and giving rise to the exchange coupling $2J_0$. (b) Chemical paths connecting coppers types A and B in neighbor chains, exchange coupled by $2J_1$. (c) Chemical paths connecting coppers pairs type A (or B) in neighbor chains, exchange coupled by $2J_2$.}
\label{fig:1}
\end{figure}

The structure is monoclinic, space group C$2/c$ (\# 15) with $a = 12.087$ \AA{}, $b = 9.583$ \AA{}, $c = 8.973$ \AA{}, $\beta = 110.85^\circ$, and $Z = 4$ molecules [Cu(NH$_2$CHCH$_3$CO$_2$)$_2\cdot$H$_2$O] in the unit cell~\cite{Calvo1991, Hitchman1987}. Two of these molecules (called A and B) are related by a rotation of $180^\circ$ around the $\bm{b}$-axis. The other two are related to the first ones by a translation within the unit cell. The copper ions are in an elongated centrosymmetric nearly octahedral coordination, equatorially bonded to two amino nitrogens and two carboxylate oxygens of \texttt{D} and \texttt{L} amino acid molecules, and apically bonded to two oxygens O$_\text{w}$ from water molecules. Alternate rotated copper ions types A and B at 4.487 \AA{} are arranged in chains along the $\bm{c}$-axis (see Fig.~\ref{fig:1}a), sharing the O$_\text{w}$ ligands providing an intrachain exchange path Cu$_\text{A}$---O$_\text{w}$---Cu$_\text{B}$ with total bond length 5.368 \AA{} and angle of $113.4^\circ$. Contribute to this bridge two ``moderate''~\cite{Jeffrey1997, Steiner2002} symmetry-related H-bonds Cu$_\text{A}$---N$_\text{eq}$---H$\cdots$O$_\text{eq}$---Cu$_\text{B}$ with $d$(N---O) = 3.015 \AA{}, typical of amino acid and protein structures~\cite{Fleck2014}, and total path-length 6.982 \AA{}, connecting equatorial N and O ligands to coppers, similar to superexchange paths studied before in other copper amino acid compounds~\cite{Siqueira1993, Rapp1995, Levstein1990a}. In comparison, Muhonen~\cite{Muhonen1986} and Talukder \textit{et al.}~\cite{Talukder2006} reported the structures and magnetic properties of the dinuclear compounds [Cu$_2$(C$_4$H$_{10}$NO)$_2$(C$_4$H$_{11}$NO)$_2$(H$_2$O)](ClC$_7$H$_4$O$_2$)$_2\cdot$2C$_3$ H$_7$OH and [Cu$_2$($\mu_2$-H$_2$O)L$_2$(H$_2$O)$_2$](ClO$_4$)$_2\cdot$2H$_2$O having pairs of copper ions bridged by water molecules and two symmetry related H-bonds as neighbor copper ions in the chains of Cu(\texttt{D},\texttt{L}-ala)$_2$ (see later). Uekusa \textit{et al.}~\cite{Uekusa1995} reported other dinuclear Cu$^\text{II}$ compound bridged by a water molecule; however, in this case also contribute two equatorial carboxylate bridges that introduce deep differences in the magnetic behavior.

\begin{figure}[tbp]
\begin{center}
\includegraphics[width=0.5\textwidth]{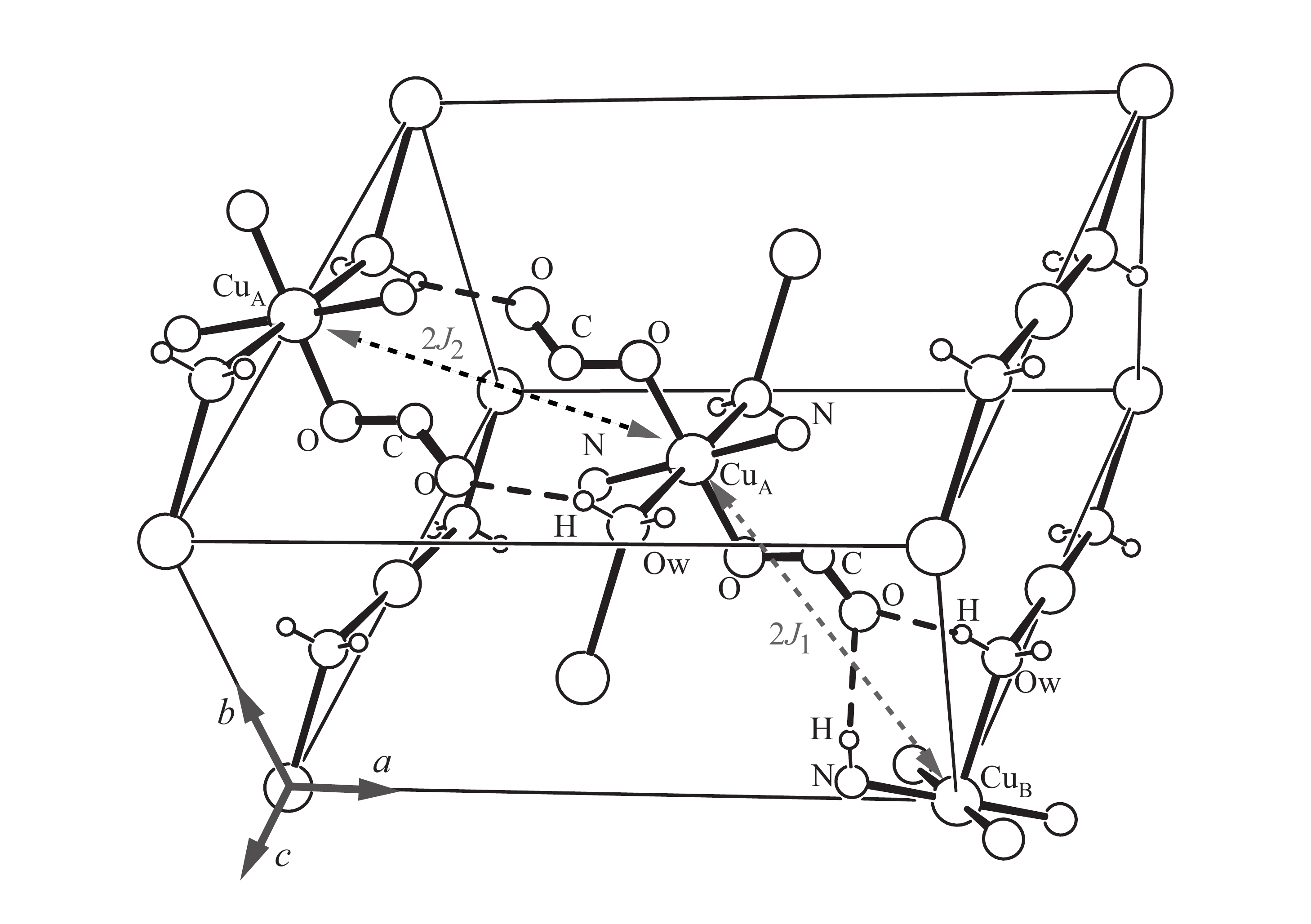}
\end{center}
\caption{Unit cell of Cu(\texttt{D},\texttt{L}-ala)$_2\cdot$H$_2$O, with the spin chains along the $\bm{c}$-axis: Interactions between coppers in neighbor chains. A central copper chain along the $\bm{c}$-axis is surrounded by four identical chains. Each copper type A interacts with four coppers type B and four type A in neighbor chains. The interchain paths types AA and AB, already described in Figs.~\ref{fig:1}b and c, are shown.}
\label{fig:2}
\end{figure}

In Cu(\texttt{D},\texttt{L}-ala)$_2$ all chains are related by translations (so, EPR measurements do not provide information about interchain interactions)~\cite{Calvo2007} and each one is surrounded by other four, at equal distances 7.712 \AA{} (Fig.~\ref{fig:2}). A copper ion type A (in the center of the unit cell in Fig.~\ref{fig:2}) has four type A copper nearest neighbors at 7.712 \AA{}, and eight coppers type B in the vertices of the monoclinic unit cell, in neighbor chains. Four AB pairs are at 7.766 \AA{}, and the other four at 9.945 \AA{}; we consider only the interactions between coppers at the shorter distance, connected by a chemical path Cu$_\text{A}$---O$_\text{eq}$---C---O$\cdots$H---N$_\text{eq}$---Cu$_\text{B}$, with five diamagnetic atoms and total path length of 9.508 \AA{}, including a H-bond with length 3.062 \AA{} and a carboxylate bridge O$_\text{eq}$---C---O, and involve equatorial O and N ligands to copper ions in neighbor chains (see Fig.~\ref{fig:1}b). Coppers of the same type (AA or BB) belonging to neighbor chains are connected by a path with two symmetry related branches of five diamagnetic atoms Cu$_\text{A}$---O$_\text{eq}$---C---O$\cdots$H---O$_\text{w}$---Cu$_\text{A}$ (total path length = 9.943 \AA{}), each including a hydrogen bond connecting an apical O$_\text{w}$ and a carboxylate bridge (see Fig.~\ref{fig:1}c). Fig.~\ref{fig:2} details one interchain path of each type, AB and AA, supporting superexchange couplings $2J_1$ and $2J_2$, respectively, that because of their length and complexity would be expected to have magnitudes much smaller than the intrachain $2J_0$, and to be responsible of the transition to the 3D magnetically ordered phase. Comparing the equatorial-equatorial character of the AB interchain connections ($2J_1$) with the equatorial-apical AA and BB connections ($2J_2$), we may assume $|2J_1| \gg |2J_2|$ and discard $2J_2$ as interchain couplings between neighbor chains~\cite{Levstein1990, Bleaney1952}.

\section{Specific heat and static magnetization data}

\begin{figure}[tbp]
\begin{center}
\includegraphics[width=0.45\textwidth]{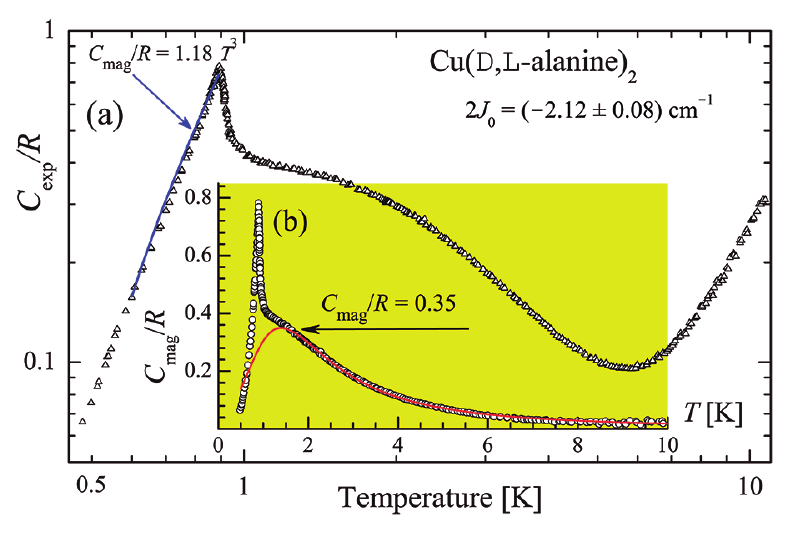}
\end{center}
\caption{(a) Log-log plot of the observed molar specific heat $C_\text{exp}/R$ vs. $T$ of Cu(\texttt{D},\texttt{L}-ala)$_2$ indicated by triangles, emphasizing the lattice contribution at high $T$. The solid blue line between 0.6 and 0.89 K indicates the observed $T^3$ dependence below the transition at $T = 0.89$ K. (b) Linear plot of the magnetic contribution to the specific heat obtained by subtracting the lattice contribution to the
experimental values. The red line is calculated for a uniform spin chain as described in the text. The horizontal arrow in (b) shows the calculated maximum value of $C_\text{mag}/R$.}
\label{fig:3}
\end{figure}

Figure~\ref{fig:3}a displays in a log-log scale the observed molar specific heat, $C_\text{exp}(T)/R$ ($R=N_\text{Av}k_B = 8.3145$ J/mol K) vs. $T$ in the temperature range $0.48 < T < 11$ K after subtracting the contributions of the sample holder and calorimeter. Values above 11 K follow equal trends, do not add new information, and are not shown. The $T$ dependence of the specific heat of Cu(\texttt{D},\texttt{L}-ala)$_2$ distinguishes different regimes. For $T > 7$ K, $\log(C_\text{exp}(T)/R)$ increases linearly with $\log(T)$ and we extrapolate $C_\text{latt}(T)/R = 234 (T/\Theta_\text{D})^3 = 2.58\times 10^{-4} T^3$, as predicted for the lattice vibrational contribution~\cite{Kittel2005} with a Debye temperature $\Theta_\text{D} \approx 97$ K, and assuming that at high $T$ the asymptotic behavior of the magnetic contribution is $C_\text{mag}\propto T^{-2}$~\cite{Carlin1986}. For $T < 7$ K $C_\text{latt}$ is small, and Fig.~\ref{fig:3}b shows that the $C_\text{mag}(T)/R = (C_\text{exp}-C_\text{latt})/R$ vs. $T$ curve displays a broad peak at $T_\text{max} = 1.8$ K, and a narrow peak at $T_N = 0.89$ K. Below 0.89 K, see Figs.~\ref{fig:3}a and b, the magnetic contribution to the specific heat follows $C_\text{mag}/R \approx 1.18 T^3$ down to 0.55 K where small deviations start to occur. The nature of the peaks in $C_\text{mag}(T)$ (see Fig.~\ref{fig:3}b) is understood by calculating the magnetic entropy $S_\text{mag}/R$ for $T\rightarrow\infty$. We estimate first $S_\text{mag}(0.48\text{ K}) = 0.0576 R$, and then we integrate numerically the experimental result in Fig.~\ref{fig:3}b such that:
\begin{equation}
S_\text{mag}(T\rightarrow\infty) = S_\text{mag}(0.48\text{ K}) + \int_{0.48\text{ K}}^{12\text{ K}} \frac{C_\text{mag}(T)}{T} dT,
\end{equation}
where $C_\text{mag}(T) \approx 0$ for $T \ge 12$ K, and $S_\text{mag}(\infty)/R \approx 0.683$ approaches asymptotically the value $\ln(2) = 0.693$ expected for the spin $1/2$ system. Thus, considering the structural information~\cite{Calvo1991, Hitchman1987}, the data in Fig.~\ref{fig:3}b indicates that the broad peak of $C_\text{mag}$ at 1.8 K arises from short range ordering within spin chains antiferromagnetically coupled by $2J_0$ and the narrow peak at $T_N = 0.89$ K marks the transition to a 3D magnetically ordered phase, where spin correlation between the chains is reached. This transition would be related to the exchange couplings $2J_1$ between Cu$^\text{II}$ ions in neighbor chains.
 
Figure~\ref{fig:4}a displays molecular magnetization ($M_z/N_\text{Av}$) isotherms of the powder sample of Cu(\texttt{D},\texttt{L}-ala)$_2$ at $T = 2$, 4 and 7 K, between 0 and 9 T that show saturation effects at high fields. Fig.~\ref{fig:4}b displays $M_z/N_\text{Av}$ for various fixed values of $B_0$ and $2 \leqslant T \leqslant 85$ K.

\section{Theory and analysis}

We describe first the thermal and magnetic properties of isolated spin chains using the method of Bonner and Fisher~\cite{Bonner1964} (BF) and calculate the intra-chain exchange couplings. Later, we estimate the exchange couplings between Cu$^\text{II}$ ions in neighbor chains, considering the observed magnetic phase transition at 0.89 K and following different approximate theories (mean field, spin waves and others).

\subsection{Spin-chains thermodynamics}

\begin{figure}[h]
\includegraphics[width=0.48\textwidth]{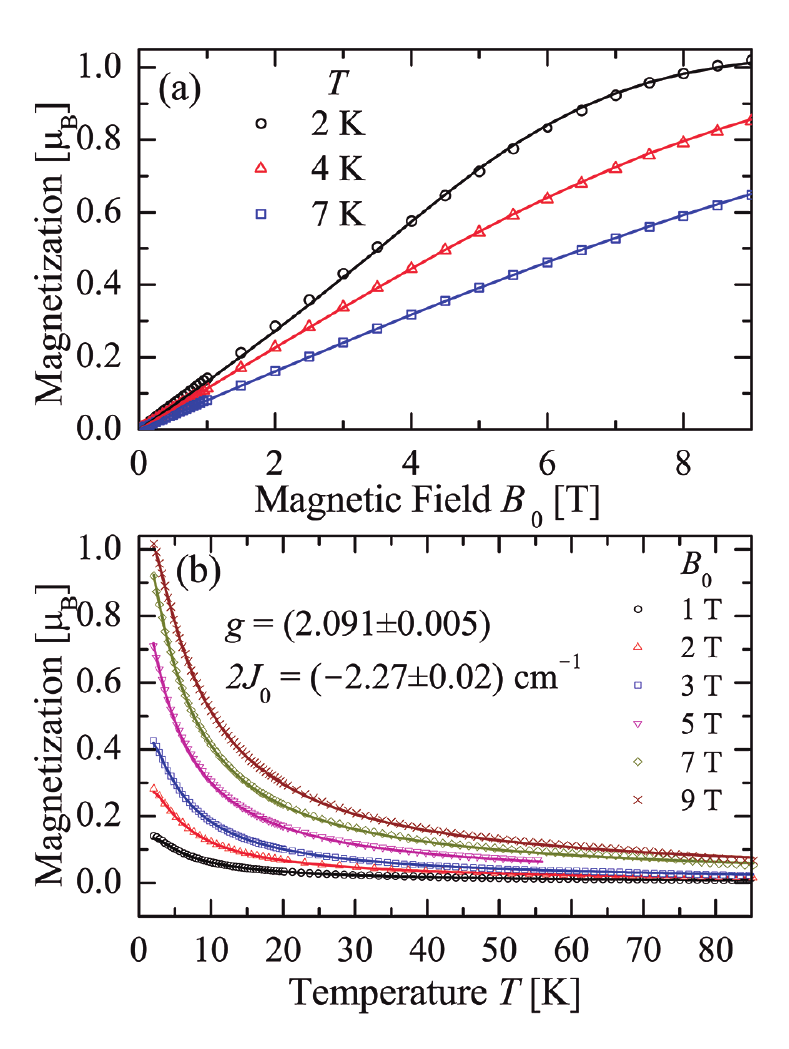}
\caption{Molecular magnetization in Bohr magnetons per Cu$^\text{II}$ spin. (a) Isothermal magnetization curves. (b) Temperature dependence of the magnetization for different values of the applied field $B_0$. The parameters $g$ and $2J_0$ given in (b) were obtained from a global least squares fit of the magnetization data. Symbols in (a) and (b) are experimental values and the lines are obtained from the fit. In all curves the agreement is within the experimental accuracy.}
\label{fig:4}
\end{figure}

A finite uniform spin-chain with $N_S$ spins having isotropic exchange coupling $2J_0$ between nearest neighbor spins and isotropic $g$-factor, under a static field $B_0$, is described by the sum of exchange and Zeeman interactions~\cite{Bonner1964, Griffiths1964},
\begin{equation}
\mathcal{H}_\text{ch}^{N_S} = |J_0| \sum_{i=1}^{N_S} \left[ -\frac{2J_0}{|J_0|}  \bm{S}_i \cdot \bm{S}_{i+1} + y S_i^z \right]
= |J_0| H_\text{ch}^{N_S},
\label{eq:ham}
\end{equation}
where $y=g\mu_B B_0/|J_0|$ is the reduced magnetic field and $S_i^z$ the $z$-component of the $i$-copper spin in the chain, and we use periodic conditions (\textit{i.e.}, spin rings with $\bm{S}_{N_S+1} = \bm{S}_1$)~\cite{Bonner1964, Chagas2006}. In powder samples, the $g$-factor is the angular average of the $g$-matrix of the Cu$^\text{II}$ ions. The molar specific heat $C_\text{mag}$, and molecular magnetization $M_z/N_\text{Av}$ for finite chains of $N_S$ spins may be calculated as a function of $T$ and $B_0$ with the BF's method~\cite{Bonner1964} using Eq.~\ref{eq:ham}, and the eigenstates of the chains through~\cite{Bonner1964, Chagas2006, Fabricius1991}:
\begin{eqnarray}
C_\text{mag}(T)   &=& \frac{R}{N_S x^2} \left[ \braket{\left(H_\text{ch}^{N_S}\right)^2} - \braket{H_\text{ch}^{N_S}}^2 \right], \label{eq:cmag}\\
\frac{M_z(B_0,T)}{N_\text{Av}} &=& \frac{g\mu_B}{N_S}\braket{S_z}, \label{eq:mmag}
\end{eqnarray}
where $x = k_B T/|J_0|$ is the ``reduced temperature.'' The angle brackets in Eqs.~\ref{eq:cmag} and \ref{eq:mmag} are thermal averages of the enclosed operators for chains with $N_S$ spins, at $T$ and $B_0$ (for $C_\text{mag}$ , $B_0 = 0$)~\cite{Chagas2006, Costa-Filho2001}:
\begin{eqnarray}
\braket{H_\text{ch}^{N_S}} &=& \frac{1}{\mathcal{Z}} \sum_{k=1}^M E_k \exp{(-J_0 E_k/k_B T)}, \\
\braket{S_z}   &=& \frac{1}{\mathcal{Z}} \sum_{k=1}^M s_k^z \exp{(-J_0 E_k/k_B T)},
\end{eqnarray}
where $\mathcal{Z}=\sum_{k=1}^M \exp{(-J_0 E_k/k_B T)}$, and $k$ labels the $M = 2^{N_S}$ states of the chain having a spin $z$-component $s_k^z$. The energy $E_k$ of the $k$-state of the Hamiltonian $H_\text{ch}^{N_S}$ of Eq.~\ref{eq:ham} is:
\begin{equation}
E_k = \epsilon_{k,0} + y s_k^z,
\end{equation}
where $\epsilon_{k,0}$ are the energies of the states in the absence of $B_0$ and in units of $|J_0|$. The thermodynamic variables for infinite chains are obtained by extrapolating to $N_S \rightarrow \infty$ the results for finite chains~\cite{Bonner1964}. We confirmed that within the considered $T$-range, chains of 20 spins provide results as valid as infinite chains and used Eq.~\ref{eq:cmag} with their eigenvalues~\cite{Chagas2006} to fit our specific heat data above 1.8 K (Fig.~\ref{fig:3}b), obtaining $2J_0 = (-2.12 \pm 0.08)$ cm$^{-1}$ and the red solid line in this figure. Also, Eq.~\ref{eq:mmag} was fitted to the whole set of magnetization data in Fig.~\ref{fig:4}, obtaining:
\begin{equation*}
g = 2.091 \pm 0.005 \text{ and } 2J_0 = (-2.27 \pm 0.02)\text{ cm}^{-1}.
\end{equation*}

The specific heat and magnetization curves calculated with these parameters (solid lines) agree well with the experiments. However, setting a lower limit in $T$ for the points considered in the fit of the specific heat avoiding the region where the 1D assumption fails, introduces an uncertainty, and we are more confident of the slightly higher value of $2J_0$ obtained from the magnetization data.

\subsection{Coupled Spin-Chains and Phase Transitions Under Different Approaches}

In the structural analysis we proposed that each copper spin type A (B) in a chain is weakly coupled to one B (A) type copper spin in four neighbor chains. Smaller interchain couplings may be neglected in a first approximation. As pointed out by Giamarchi~\cite{Giamarchi2003}, one expects that since the individual chains are AFM, interchain interactions $2J_1$ of either sign would stabilize this AFM condition in the 3D array. The transition to a 3D ordered phase (see Fig.~\ref{fig:3}) cannot be treated exactly~\cite{Giamarchi2003} and many theoretical~\cite{Ohmae1995, Matsumoto2000, Sakakibara2002, Oguchi1964, Smart1966, Schulz1996, Irkhin2000} and numerical~\cite{Yasuda2005} efforts have been made to deal with the problem. According to the spin-wave (sw) theory, the magnetic contribution $C_\text{mag}(T)$ below the transition temperature $T_N$ should behave as~\cite{DeJongh1974, Ohmae1995, Matsumoto2000, Sakakibara2002}:
\begin{equation}
C_\text{mag}(\text{sw}) \propto T^{d/n}, 
\end{equation}
where $d = 3$ is the magnetic dimension at low $T$ and $n = 1$ for AFM and $n = 2$ for FM couplings. So, the experimental result $C_\text{mag}/R \propto T^3$ below $T_N$ reflects an AFM 3D ordering below 0.89 K. Small deviations at the lowest $T$ are justified because, besides $2J_1$, other non-negligible interchain interactions may contribute (see Figs.~\ref{fig:1} and \ref{fig:2}), and because of the low lattice space symmetry. We first estimated $2J_1$ using the sw results of Sorai \textit{et al.}~\cite{Ohmae1995, Matsumoto2000, Sakakibara2002}, considering axial symmetry around the chain axis, with exchange coupling $2J_0$ along the spin chains, and $2J_1$ between chains, using~\cite{Ohmae1995}:
\begin{equation}
\frac{C_\text{mag}(\text{sw})}{R} = \frac{\zeta(4)\Gamma(4)k_B^3}{16 \pi^2 S^3 V} T^3,
\end{equation}
and comparing with the experimental result, $C_\text{mag}/R \approx 1.18 T^3$. $\zeta(4) = \pi^4/90$ is Riemann's zeta function and $\Gamma(4) = 3!$ is Euler's gamma function, and~\cite{Ohmae1995},
\begin{equation}
V = 
\begin{cases} 
(|J_0|+2|J_1|)^{3/2} |J_0J_1^2|^{1/2} &\mbox{for } J_1 < 0 \\
|J_0^2J_1| &\mbox{for } J_1 > 0.
\end{cases}
\end{equation}
So, for $2J_0 = -2.27$ cm$^{-1}$ and $J_1 < 0$, it is $2J_1 = -0.125$ cm$^{-1}$, while for $J_1 > 0$, it is $2J_1 = 0.146$ cm$^{-1}$. Meanwhile, assuming axial symmetry around the chain axis and defining $\eta = |J_1/J_0|$, the mean field (mf) theory of Oguchi gives:~\cite{Oguchi1964}
\begin{equation}
\frac{k_B T_N}{|J_0|} = \frac{4}{3}\frac{S(S+1)}{I(\eta)}
\label{eq:mf1}
\end{equation}
where~\cite{Oguchi1964, Montroll1956, Hennessy1973}, 
\begin{equation}
I(\eta) \approx 0.64 \eta^{-1/2}
\label{eq:mf2}
\end{equation}
Equations \ref{eq:mf1} and \ref{eq:mf2} allow estimating the ratio $\eta = 0.122$ from the experimental values of $T_N$ and $2J_0$ and so $|2J_1| = 0.28$ cm$^{-1}$. In addition to these sw and mf estimates for the interchain coupling, we now turn to more recent analyses for the transition temperature in low-dimensional systems based on a mean-field theory within a random-phase approximation (RPA)~\cite{Schulz1996, Irkhin2000}. According Schulz~\cite{Schulz1996}, the mean-field treatment of the full 3D Hamiltonian yields for the interchain coupling within RPA: 
\begin{equation}
\eta = \frac{k_B T_N/|2J_0|}{4c \ln^{1/2} \left( \dfrac{\lambda |2J_0|}{k_B T_N} \right)},
\label{eq:schulz}
\end{equation}
where $c = 0.32$ and $\lambda = 5.8$ were estimated from numerical calculations in a single chain~\cite{Starykh1997}. Therefore, $\eta = 0.122$ and hence we obtain the same $2J_1$ as Oguchi's estimate. The agreement between the two estimates is, however, accidental, since these approximations predict different scalings for the transition temperature. A modified RPA approach was reported by Irkhin and Katanin~\cite{Irkhin2000} in which the exchange ratio obeys:
\begin{equation}
\eta = \frac{k_B T_N/|2J_0|}{4c \sqrt{\ln \left( \dfrac{\lambda |2J_0|}{k_B T_N} \right) + \dfrac{1}{2} \ln \ln \left( \dfrac{\lambda |2J_0|}{k_B T_N} \right)} },
\label{eq:irkhin}
\end{equation}
where $c = 0.23$ includes a correction factor $\sim$0.7 accounting for spin fluctuation along the directions perpendicular to the chains.~\cite{Irkhin2000} In this case we obtain $|2J_1| = 0.36$ cm$^{-1}$. The same expression as Eq.~\ref{eq:irkhin} is given ``empirically'' by Yasuda \textit{et al.}~\cite{Yasuda2005}, where a numerical fit based in quantum Monte Carlo (QMC) simulations yields $c = 0.233$ and $\lambda = 2.6$ and the interchain interaction in Cu(\texttt{D},\texttt{L}-ala)$_2$ results $|2J_1| = 0.41$ cm$^{-1}$.

\section{Discussion}

In this work we studied the specific heat and magnetization for the copper amino acid salt Cu(\texttt{D},\texttt{L}-ala)$_2$ in a wide temperature and magnetic field range. The compound displays a clear 1D spin chain behavior above $\sim$1.8 K, that was observed in a previous work based in basic susceptibility data~\cite{Calvo1991}. Our present data allows a precise characterization of this spin chain behavior and adds up to detect a 3D magnetic phase transition at 0.89 K that was analyzed under different approximations in order to evaluate the exchange interactions $2J_1$ between coppers in neighbor chains. Here we compare the results of these estimations. We also analyze the values of $2J_0$ and $2J_1$ in terms of the chemical paths connecting the copper ions and the roles of H-bonds in supporting exchange couplings.

\subsection{Discussion of the magnitudes of the interchain interactions}

Since no direct evaluation of the interchain coupling is available for Cu(\texttt{D},\texttt{L}-ala)$_2$, the comparison between the estimated values of $2J_1$ is indeed subjected to the assumptions made in the considered approximations. On one hand, the mf estimations exhibit a systematic increasing of the ratio $\eta = |J_1/J_0|$ in Oguchi~\cite{Oguchi1964} and Schulz~\cite{Schulz1996} ($\eta = 0.12$) and Irkhin and Katanin~\cite{Irkhin2000} ($\eta = 0.16$), as compared with the QMC result~\cite{Yasuda2005} $\eta = 0.18$, that can be attributed to an overestimated critical temperature $T_N$, since in these approximations thermal and quantum fluctuations, which tend to inhibit the phase transition, are taken into account perturbatively. On the other hand, the spin-wave approach~\cite{Ohmae1995, Matsumoto2000, Sakakibara2002} estimates a smaller ratio $\eta = 0.06$, which may indicate failures in the collective-mode description near the critical temperature~\cite{Oguchi1963}.

The rather large value of the ratio $\eta$, due to a transition temperature $k_B T_N = 0.27 |2J_0|$, suggests that the chains are not as well isolated as in other compounds (see, \textit{e.g.} Ref.~\cite{Blundell2007}). This was surprising considering the relatively long chemical path connecting neighbor chains as compared with those involved in the intrachain exchange coupling. An explanation is that the intervening carboxylate bridge in the interchain chemical path would act as a ``short circuit'', leaving the H-bond in series as the dominant contribution to $2J_1$. 

It is important to notice that the above estimates are based on models where the underlying lattice structure is highly symmetric, while in Cu(\texttt{D},\texttt{L}-ala)$_2$ these symmetries are clearly absent. In particular, the studied effective coupling between neighbor chains in the compound's monoclinic structure is not strictly perpendicular to the direction of the chains, as can be seen in Fig.~\ref{fig:2}, and thereby the above invoked axial symmetry is accomplished only partially. Besides, other interchain couplings like the aforementioned $2J_2$ may also introduce deviations with respect to the simplified approximations, where a single interchain exchange path is accounted for. With this unclear situation we opted to report a wide range $0.1 \leqslant |2J_1| \leqslant 0.4$ cm$^{-1}$ for the interchain interactions, and analyze this result below.

\subsection{Magnetostructural correlations of intrachain and interchain interactions}

As shown by Figs.~\ref{fig:1} and \ref{fig:2}, the structure~\cite{Calvo1991, Hitchman1987} of Cu(\texttt{D},\texttt{L}-ala)$_2$ clearly indicates that the observed 1D magnetic behavior arises from chains along $\bm{c}$, where pairs of neighbor coppers ions at 4.487 \AA{} share a water apical oxygen atom. Also contribute to this exchange path two symmetry-related H-bonds with total path length 6.982 \AA{} connecting N and O equatorial ligands to copper ions and N---O lengths 3.015 \AA{}. This AFM superexchange interaction $2J_0 = -2.27$ cm$^{-1}$ arises from a quantum process described by several authors~\cite{Kahn1993, Anderson1959, Hay1975, Kahn1985}, and cases where more than one bond contribute have been treated by Levstein \textit{et al.}~\cite{Levstein1990b}, who discussed rules to tune these paths in order to maximize the coupling. We analyzed the relative role of the water oxygens and the H-bonds and propose that the water oxygen paths in Cu(\texttt{D},\texttt{L}-ala)$_2$ (apical ligand to neighbor coppers along $\bm{c}$) do not play an important role in the coupling because the overlap between the involved magnetic orbitals is expected to be very small. We further support this assumption considering that these are ``moderate'' H-bonds~\cite{Jeffrey1997, Steiner2002} and their individual contributions add up due to symmetry conditions~\cite{Levstein1990b}. Assuming that the main contribution to the interaction $2J_0$ arises from the symmetry related H-bonds described before is in line with the arguments given by Muhonen~\cite{Muhonen1986} and Talukder \textit{et al.}~\cite{Talukder2006} The much larger exchange couplings reported in these cases may be attributed to much shorter and stronger H-bonds with total lengths $d = 2.45$ and 2.63 \AA{}, respectively. The results of the DFT calculations of $2J_0$ of Talukder \textit{et al.}~\cite{Talukder2006}, including and neglecting the water oxygen in the exchange path, also support our assumption.
We also consider the results for the complexes of copper with the \texttt{L}-aminobutyric acid (\texttt{L}-but) and with its racemic mixture (\texttt{D},\texttt{L}-but)~\cite{Siqueira1993, Levstein1990a}, where H-bonds with properties similar to those along $\bm{c}$ in Cu(\texttt{D},\texttt{L}-ala)$_2$ support AFM exchange interactions $2J_0 = -1.28$ and $-1.68$ cm$^{-1}$ in Cu(\texttt{L}-but)$_2$ and Cu(\texttt{D},\texttt{L}-but)$_2$, respectively. In these cases the two H-bonds are slightly different in Cu(\texttt{L}-but)$_2$ but symmetry related in Cu(\texttt{D},\texttt{L}-but)$_2$, with total bond lengths $7.086$ and $6.946$ \AA{}, respectively, and the spin chains are well isolated with no phase transitions observed in the specific heat above 0.070 K~\cite{Siqueira1993}.
 
Regarding to the coupling $2J_1$ between chains giving rise to the phase transition, we assign it to superexchange paths connecting coppers at $7.766$ \AA{} through Cu$_\text{A}$---O$_\text{eq}$---C---O$\cdots$H---N$_\text{eq}$---Cu$_\text{B}$ bridges containing a moderate H-bond in series with a carboxylate bridge, connecting N and O equatorial ligands of Cu$^\text{II}$ in neighbor chains, displayed in Figs.~\ref{fig:1}b and \ref{fig:2}. Considering that the exchange lengths vary from $6.982$ \AA{} in the intrachain path, to $9.508$ \AA{} in the interchain path, and the number of diamagnetic atoms in the paths vary from 3 to 5, the relatively close values of $2J_1$ and $2J_0$ suggest that the \textit{covalent} carboxylate bridge in $2J_1$ introduces a small effect in the overall magnitude of the interaction so the contribution from the H-bond would be the dominant one. 

\section{Conclusions}

The specific heat and magnetization of the compound Cu(\texttt{D},\texttt{L}-ala)$_2$ display above $T = 1.8$ K, a well-defined 1D AFM behavior with $2J_0 = (-2.27\pm 0.02)$ cm$^{-1}$ and $g = 2.091 \pm 0.005$. Considering the crystal structure the exchange interaction between neighbor coppers is undoubtedly supported by a chemical path containing an apical water oxygen ligand and two symmetry related H-bonds connecting equatorial oxygen and nitrogen ligands to copper. However, we attribute the main contribution to this interaction to the H-bonds, neglecting the effect of the water molecule. 
The specific heat data displays a transition to AFM 3D order at $T_N = 0.89$ K. This result is used to estimate under different approximations a wide range $0.1 \leqslant |2J_1| \leqslant 0.4$ cm$^{-1}$ for the exchange interaction coupling spins in neighbor chains that are supported by paths with total length of $9.5$ \AA{}. The relatively close magnitudes of inter- and intrachain couplings may suggest that the carboxylate bridge acts as a ``short circuit'' and has little influence in $2J_1$, leaving the H-bond as its principal contributor.

\section*{Acknowledgements}
Work supported by program CAI+D-UNL in Argentina and by CNPq, FAPERJ and CAPES in Brazil. RC and HLC are members of CONICET. We acknowledge help from Dr. Daniel Rodrigues in early stages of this work.\\

\bibliographystyle{elsarticle-num}
\biboptions{sort&compress}
\bibliography{cite}

\end{document}